\documentclass[aps, prx, twocolumn, superscriptaddress, amsmath, amssymb]{revtex4-2}

\usepackage{amsmath,amssymb,amsfonts,physics}    
\usepackage{graphicx}      
\usepackage{fancyhdr}    
\usepackage{xcolor}
\usepackage{tcolorbox} 
\usepackage{enumitem}  
\usepackage{titlesec}  
\usepackage[colorlinks=true,
linkcolor=blue,
urlcolor=blue,
citecolor=blue,
anchorcolor=blue]{hyperref}

\newcommand{\nn}{\nonumber \\}
\newcommand{\su}{\uparrow}
\renewcommand{\sd}{\downarrow}
\newcommand{\dg}{^{\dagger}}

\begin{document}

\title{Quantum capacitance and parity switching of a quantum-dot-based Kitaev chain}
\author{Chun-Xiao Liu}
\email{contact author: chunxiaoliu62@sjtu.edu.cn}
\affiliation{Tsung-Dao Lee Institute, Shanghai Jiao Tong University, Shanghai 201210, China}
\affiliation{School of Physics and Astronomy, Shanghai Jiao Tong University, Shanghai 200240, China}
\affiliation{State Key Laboratory of Micro and Nano Engineering Science, Shanghai Jiao Tong University, Shanghai 201210, China}

\date{\today}

\begin{abstract}
An array of quantum dots coupled via superconductivity provides a new platform for creating Kitaev chains with Majorana zero modes, offering a promising avenue toward topological quantum computing.
In this work, we theoretically study the quantum capacitance of a minimal Kitaev chain weakly coupled to an external normal lead.
We find that in the open regime, charge stability diagrams of quantum capcaitance can help to identify the sweet spot of a Kitaev chain, consistent with tunnel spectroscopy.
Moreover, the quantum capacitance of a single quantum dot coupled to Andreev bound states reveals the interplay between two distinct parity switching mechanisms: coupling to an external normal lead and intrinsic quasiparticle poisoning.
Our work provides useful physical insights into the quantum capacitance and parity dynamics in a quantum-dot-based Kitaev chain device.
\end{abstract}

\maketitle

\section{Introduction}


Quantum dot-superconductor hybrids provide a versatile platform for realizing quantum devices.
In the field of quantum information processing, a quantum dot embedded in a Josephson junction forms an Andreev spin qubit~\cite{Zazunov2003Andreev,Chtchelkatchev2003Andreev,Wendin2007Quantum,Padurariu2010Theoretical,Park2017Andreev,PitaVidal2023Direct}, and superconducvitiy connecting two distant spin qubits can implement long-range gate operations~\cite{Leijnse2013Coupling, Spethmann2024High}.
Double quantum dots connected by a superconductor have been used as a Cooper pair splitter to study quantum entanglement in solid states~\cite{Beckmann2004Evidence,Hofstetter2009Cooper,Das2012High,Tan2015Cooper, Wang2022Singlet,Wang2023Triplet}.
In addition, an array of alternating quantum dot and superconductivity enables the engineering of topological Kitaev chains with non-Abelian Majorana zero modes~\cite{Sau2012Realizing,Leijnse2012Parity,Fulga2013Adaptive}, which are the building blocks of topological quantum computing~\cite{Kitaev2001Unpaired, Nayak2008Non-Abelian, Alicea2012New, DasSarma2015Majorana}.

Over the past few years, significant experimental progress has been made in realizing two- and three-site Kitaev chains in coupled quantum dots, supported by observations of robust zero-bias conductance peaks as evidence of Majorana zero modes~\cite{Dvir2023Realization, tenHaaf2024Twosite,Bordin2024Crossed, Zatelli2024Robust,Bordin2025Enhanced, tenHaaf2025Observation,Bordin2025Probing, tenHaaf2025Probing}.
Crucially, here search of the Majorana sweet spot is achieved by controlling the ratio of elastic cotunneling and crossed Andreev reflection mediated by the Andreev bound states (ABSs) in the hybrid region~\cite{Liu2022Tunable,Bordin2023Tunable}.
In parallel, extensive theoretical researches have been performed on such systems~\cite{Tsintzis2022Creating,Souto2023Probing, Liu2024Enhancing, Miles2024Kitaev, Luna2024Flux, Luethi2024From, Liu2024Coupling, Ezawa2024Even_odd, Dourado2025Majorana, Pandey2025Crystalline, Bozkurt2025Interaction, Liu2025Scaling, Zhang2025Poor}.
All these efforts are paving the way toward the next milestones in this field: demonstration of the non-Abelian statistics of Majorana anyons and implemention of Majorana qubits~\cite{Liu2023Fusion,Boross2024Braiding,Pandey2024Nontrivial, Tsintzis2024Majorana, Miles2025Braiding, Pan2025Rabi}.

Quantum capacitance provides a particularly powerful probe of nanoscale hybrid devices~\cite{Malinowski2022Radio,deJong2021Rapid,Ibberson2021Large,Han2023Variable,Petersson2010Charge, Aghaee2025Interferometric}. 
It enables readout of the joint fermion parity of a Majorana pair~\cite{Liu2023Fusion} as well as the ground-state parity of a Majorana qubit~\cite{Pan2025Rabi}, which are the basic elements of measurement-based topological quantum computing~\cite{Bonderson2008Measurement,Plugge2017Majorana,  Karzig2017Scalable}.
Recently, single-shot readout of Majorana parity has been successfully demonstrated in a two-site Kitaev chain~\cite{vanLoo2025Single}, in agreement with theoretical predictions~\cite{Liu2023Fusion}.
In addition, gate-reflectometry techniques are applied in the measurement of a single quantum dot, further expanding the scope of quantum capacitance in Kitaev chain devices~\cite{Zhang2025Gate}.

Motivated by these experimental progresses, here we theoretically investigate the quantum capacitance of a two-site Kitaev chain which is either coupled or decoupled to a normal reservoir, focusing on the signatures of the Majorana sweet spot.
Then, we consider a single quantum dot coupled to ABSs by detuning one of the quantum dots far from resonance.
We find that now quantum capacitance can provide additional information about Hamiltonian parameters, such as dot-ABS coupling strength and coherence factors.
Moreover, it reveals the interplay between two distinct parity switching mechanisms: coupling to an external normal lead and intrinsic quasiparticle poisoning. 

\begin{figure}[tbp]
    \centering
    \includegraphics[width=\linewidth]{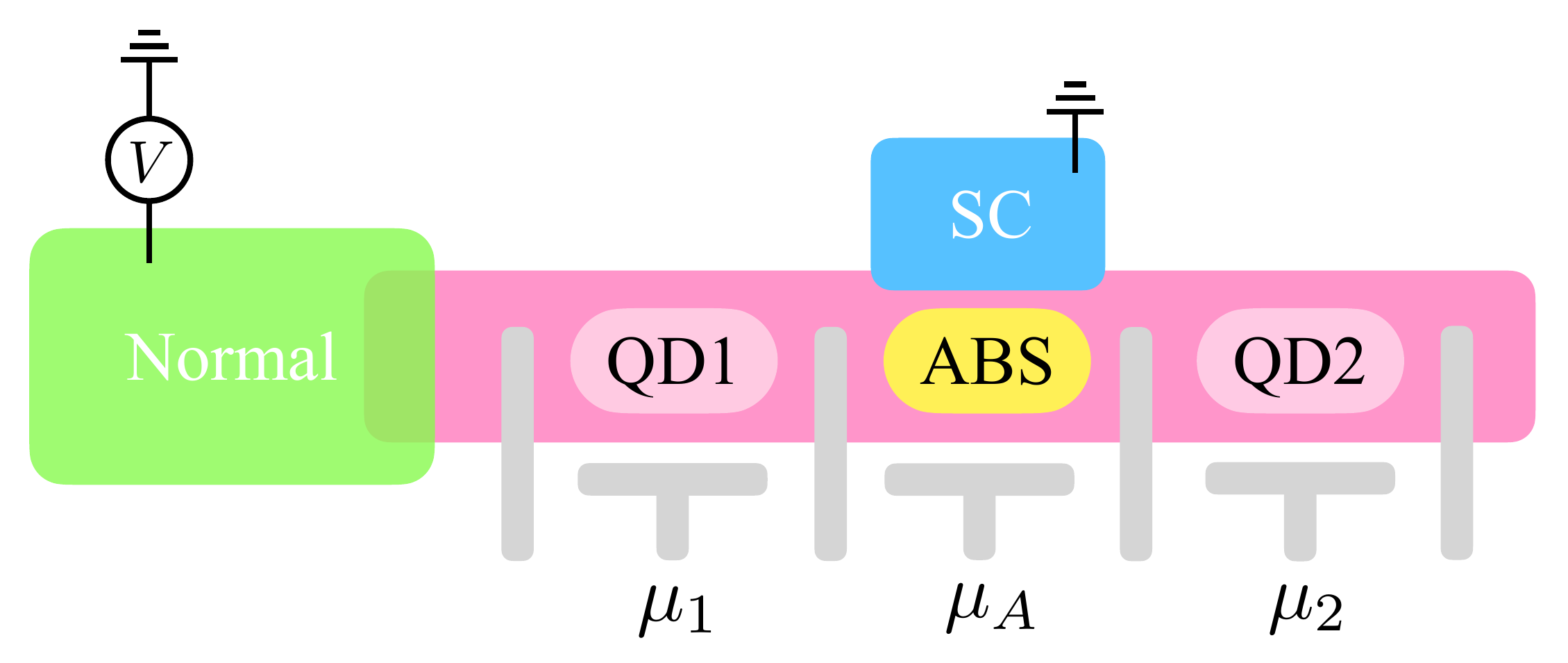}
    \caption{Schematic of a two-site Kitaev chain device. 
    Quantum dots are defined in the semiconductor using electrostatic gates (gray lines).
    They are connected via Andreev bound states in the hybrid superconducting region.
    A normal-metal lead is weakly coupled to the device from the left, with the coupling strength controlled by a tunnel gate.}
    \label{fig:schematic}
\end{figure}

\section{Model Hamiltonian and methods}
A minimal Kitaev chain device consists of two gate-defined quantum dots (QDs) connected by ABS in a hybrid superconducting region, see Fig.~\ref{fig:schematic}. 
The model Hamiltonian is given by
\begin{align}
& H_K = H_{D} + H_{A} + H_{\text{tunn}}, \nn
& H_{D} = \sum_{i=1,2} \mu_{i} n_{i\sd} + (\mu_{i} + 2E_{Zi}) n_{i\su} + U_{i}n_{i\su}n_{i\sd},\nn
& H_A = \mu_{A} (n_{A\su} + n_{A\sd}) + \Delta_0 (c\dg_{A\su}c\dg_{A\sd} + c_{A\sd} c_{A\su} ), \nn
&H_{\text{tunn}} = \sum_{\sigma=\su,\sd} \big[  t_{\text{sc}} (c\dg_{A\sigma} c_{1\sigma} + c\dg_{2\sigma} c_{A\sigma} ) \nn
& \quad +  \sigma t_{\text{sf}} ( c\dg_{A\overline{\sigma}} c_{1\sigma} + c\dg_{2\overline{\sigma}} c_{A\sigma} ) + \text{H.c.} \big].
\label{eq:H_K}
\end{align}
Here $H_D$ is the Hamiltonian for quantum dots, where $n_{i\sigma} = c\dg_{i\sigma}c_{i\sigma}$ is the occupancy of the orbital in the $i$th QD  with spin $\sigma$, $\mu_i$ is the chemical potential controlled by a plunger gate nearby, $E_{Zi}$ is the strength of the Zeeman spin splitting induced by an externally applied magnetic field, and $U_i$ is the onsite Coulomb energy. 
$H_A$ is the Hamiltonian for the ABSs formed in the hybrid superconducting region, where $\mu_A$ is the chemical potential of the normal state, and $\Delta_0$ is the $s$-wave pairing potential induced by proximity effect from a superconducting lead.
$H_{\text{tunn}}$ describes electron hopping between neighboring QDs and ABSs, with $t_{\text{sc}}$ being the amplitude for spin-conserving processes and $t_{\text{sf}}$ for spin-flipping ones due to spin-orbit interaction.
In the rest of this work, we set $\Delta_0=1$ to be the unit of energy scales.
Other fixed parameters include $E_{Z1} = E_{Z2}=3$ and $U_1=U_2=5$.
When a normal metal lead is weakly coupled to the Kitaev chain, it gives
\begin{align}
    &H_{\text{ext}} = H_{\text{lead}} + H_{\text{coupling}} \nn
    & = \sum_{k,\sigma} \left[ ( \varepsilon_k - \mu_{\text{lead}} ) n_{k\sigma} + t_c \left( c\dg_{1\sigma} f_{k\sigma} + f\dg_{k\sigma} c_{1\sigma} \right) \right],
\end{align}
where $n_{k\sigma}=f\dg_{k\sigma} f_{k\sigma}$ is the occupancy of the lead electrons with energy $\varepsilon_k=\hbar^2k^2/2m_e$ and spin $\sigma$, $\mu_{\text{lead}}$ is the chemical potential that can be varied by lead voltage, and $t_c$ is the tunneling strength between the lead and QD1, which can be controlled by a tunnel gate, see Fig.~\ref{fig:schematic}.

The presence of either an external electron reservoir or quasiparticle poisoning would drive the system out of equilibrium.
The resulting non-Fermi-Dirac distribution can be obtained by solving the semiclassical rate equation as below~\cite{Beenakker1991Theory, Nazarov2009Quantum}
\begin{align}
    d P_{\alpha} / dt = \sum_{\beta \neq \alpha} \left( \Gamma_{\alpha \beta} P_{\beta} - \Gamma_{ \beta \alpha} P_{\alpha} \right)=0, \quad \sum_{\alpha} P_{\alpha}=1,
\end{align}
where $\{P_{\alpha} \}$ is the probability distribution, $\Gamma_{\beta \alpha}$ is the rate of transition from $\ket{\alpha}$ to $\ket{\beta}$, which are the eigenstates of the Hamiltonian in Eq.~\eqref{eq:H_K}.
$d P_{\alpha} / dt=0$ denotes the steady-state condition, and $\sum_{\alpha} P_{\alpha}=1$ is the normalization condition.
Here, $\Gamma_{ \beta \alpha}$ includes three possible mechanisms.
The first mechanism is single-electron tunneling from the normal lead, which gives the following rates:
\begin{align}
    &\Gamma_{\beta \alpha, \text{SET}} = \Gamma^{+}_{\beta \alpha}+ \Gamma^{-}_{\beta \alpha}, \nn
    & \Gamma^{+}_{\beta \alpha} =\gamma_{\text{lead}} \sum_{\sigma=\su,\sd}  \abs{\bra{\beta} c\dg_{1\sigma} \ket{\alpha}}^2  n_F(\varepsilon_{\beta} -\varepsilon_{\alpha}-eV,T_{\text{lead}} ) \nn
    &\Gamma^{-}_{\beta \alpha}= \gamma_{\text{lead}} \sum_{\sigma=\su,\sd} \abs{\bra{\beta} c_{1\sigma} \ket{\alpha}}^2  \nn
    &\qquad \times [1- n_F(\varepsilon_{\alpha} -\varepsilon_{\beta}-eV, T_{\text{lead}} )],
\end{align} 
where $\gamma_{\text{lead}}=2\pi t^2_c \rho $ is the lead-dot tunneling rate with $\rho$ being the density of states at the Fermi energy of the lead, $\varepsilon_{\alpha}$ is the eigenenergy of state $\ket{\alpha}$, and $V$ is the bias voltage set in the lead relative to the grounded superconductor.
$n_F(x)$ is the Fermi-Dirac function with $T_{\text{lead}}$ being the lead temperature.
$\Gamma^{+}_{\beta \alpha}$ or $\Gamma^{-}_{\beta \alpha}$ flips the fermion parity of the Kitaev chain device by either adding or removing an electron.
The second mechanism is quasiparticle poisoning, leading to the rate as below:
\begin{align}
    \Gamma_{\beta \alpha, \text{qpp}} = \gamma_{\text{qpp}} n_F(\varepsilon_{\beta} - \varepsilon_{\alpha}, T_{\text{qpp}}) ,
    \label{eq:Gamma_qpp}
\end{align} 
where $\gamma_{\text{qpp}}$ is the characteristic poisoning rate, $T_{\text{qpp}}$ is the effective temperature.
Here, transitions are allowed only between states with opposite fermion parity.
The third mechanism is relaxation, which describes the transition from a higher-energy state to a lower one within the same parity subspace.
The relaxation rates are given by
\begin{align}
    \Gamma_{\beta \alpha, \text{relax}} = \gamma_{\text{relax}} \theta(\varepsilon_{\alpha} - \varepsilon_{\beta} ) ,
    \label{eq:Gamma_relax}
\end{align} 
where $\gamma_{\text{relax}}$ is the characteristic relaxation rate, and $\theta(x)$ is the Heaviside step function, which equals unity for $x>0$ and zero otherwise.
Once the probability distribution $\{ P_{\alpha} \}$ is obtained, the stationary current and differential conductance can be calculated as follows
\begin{align}
    G = dI/dV, \quad I = e \sum_{\alpha, \beta} (\Gamma^+_{\beta \alpha} - \Gamma^-_{\beta \alpha} ) P_{\alpha},
    \label{eq:G} 
\end{align}
while the averaged quantum capacitance is given by 
\begin{align}
    \langle C_q \rangle = \sum_{\alpha} P_{\alpha} C_{q,\alpha} = -\sum_{\alpha} P_{\alpha} \frac{d^2 \varepsilon_{\alpha}}{d\mu^2_1}.
    \label{eq:C_q}
\end{align}
Here, the quantum capacitance is defined as the second derivative with respect to QD1, corresonding to the measurements reported in Ref.~\cite{Zhang2025Gate}.
A minus sign is included in Eq.~\eqref{eq:C_q} to ensure $\expval{C_q}>0$, without loss of generality.

\begin{figure*}[tbp]
    \centering
    \includegraphics[width=0.8\linewidth]{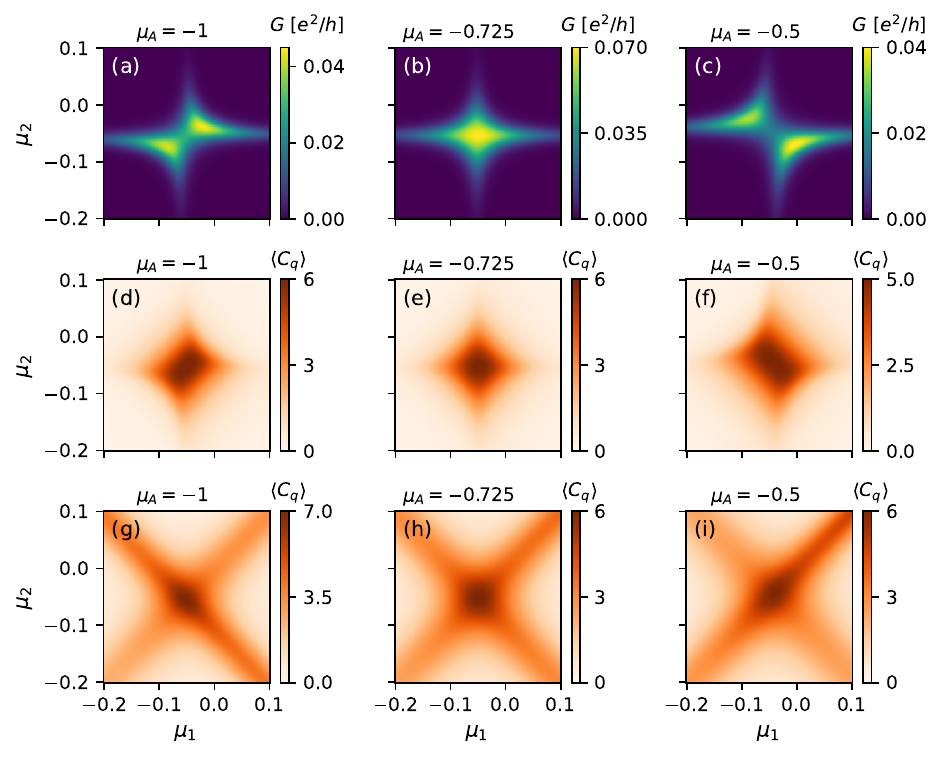}
    \caption{Conductance and quantum capacitance of a two-site Kitaev chain device.
    (a)-(c): Conductance in the open regime, assuming $\gamma_{\rm lead} =0.003, k_BT_{\rm lead}=0.005$, and $\gamma_{\rm qpp}=\gamma_{\rm relax}=0$.
    The three panels correspond to the cases of normal coupling being stronger than, equal to, or weaker than superconducting coupling.
    (d)-(f): Quantum capacitance in the open regime.
    (g)-(i): Quantum capacitance in the closed regime, assuming $\gamma_{\rm lead}=0$ and $\gamma_{\rm relax}=1 \gg \gamma_{\rm qpp} =0.001$, and $k_BT_{\rm qpp}=2$.
    Other parameters: $t_{\rm sc}=0.3, t_{\rm sf}=0.1$.}
    \label{fig:G_Cq_pmm}
\end{figure*}

\section{Two-site Kitaev chains}

We first consider the charge stability diagrams of a two-site Kitaev chain in the open regime where the tunneling rate from the normal lead is dominantly larger than the poisoning rate. 
Without loss of generality, we assume $\gamma_{\rm lead}>0$ and $\gamma_{\rm qpp}=\gamma_{\rm relax}=0$.
Figures~\ref{fig:G_Cq_pmm}(a)–\ref{fig:G_Cq_pmm}(c) show the numerically calculated differential conductance for different values of $\mu_A$ in unit of $\Delta_0$.
As $\mu_A$ varies, the anticrossing of the conductance lines reverses direction, reflecting a crossover from dominant normal coupling at $\mu_A=-1$ [see Fig.~\ref{fig:G_Cq_pmm}(a)] to dominant superconducting coupling at $\mu_A=-0.5$ [see Fig.~\ref{fig:G_Cq_pmm}(c)].
In particular, at $\mu_A=-0.725$, the charge stability diagram exhibits a crossing pattern [see Fig.~\ref{fig:G_Cq_pmm}(b)], indicating that the two types of couplings are balanced, reaching the Majorana sweet spot.
Quantum capacitance is calculated with the same parameters, as shown in Figs.~\ref{fig:G_Cq_pmm}(d)-\ref{fig:G_Cq_pmm}(f).
The key feature is the evolution of the $\langle C_q \rangle$ peak with $\mu_A$.
It extends along the $\mu_1=\mu_2$ ($\mu_1=-\mu_2$) direction when the normal (superconducting) coupling dominates, while at the sweet spot, only a small diamond-shaped peak remains near the center. 
To understand these features, we use the following effective Hamiltonian of a two-site Kitaev chain:
\begin{align}
    & H_{\text{eff}} = H_{\text{even}} \oplus H_{\text{odd}}, \nn
    & H_{\text{even}} = 
    \begin{pmatrix}
         0 & \Delta'\\
         \Delta' & \mu_1 + \mu_2
    \end{pmatrix}, \quad
    H_{\text{odd}} =
    \begin{pmatrix}
        \mu_1 & t' \\
        t' & \mu_2
    \end{pmatrix},
\end{align}
where the bases are $\{ \ket{00}, \ket{11} \}$ in the even-parity subspace and $\{ \ket{10}, \ket{01} \}$ in the odd-parity one.
$t'$ and $\Delta'$ are the effective normal and superconducting couplings mediated by the ABS.
Note that in the strong Zeeman field regime, quantm dots enter the spinless regime, and thereby only spin-down orbitals are considered near $\mu_1 \approx \mu_2 \approx 0$.
The eigenenergies and quantum capacitance are
\begin{align}
    & E_{eg} = \mu_+ - \sqrt{\mu^2_+ + \Delta'^2},\quad C_{q,eg} = \frac{\Delta'^2}{4(\mu^2_+ + \Delta'^2)^{3/2}}, \nn
    & E_{og} = \mu_+ - \sqrt{\mu^2_- + t'^2},\quad C_{q,og} = \frac{t'^2}{4(\mu^2_- + t'^2)^{3/2}},
    \label{eq:Cq_analytic}
\end{align}
where $eg$ and $og$ denote the ground states in the even- and odd-parity subspaces, and $\mu_{\pm} = (\mu_1 \pm \mu_2)/2$.
As seen from Eq.~\eqref{eq:Cq_analytic}, the $C_q$ peak extends along the $\mu_1 =\pm \mu_2$ directions for the odd- and even-parity ground states, respectively, explaining the numerical results in Figs.~\ref{fig:G_Cq_pmm}(d)-\ref{fig:G_Cq_pmm}(f).
Importantly, in the open regime, coupling to the normal lead relaxes the system to the global ground state, so that both conductance and quantum capacitance can reflect the nature of the dominant coupling between quantum dots.

On the other hand, we also consider the closed regime where the normal lead is decoupled from the device, i.e., $\gamma_{\rm lead}=0$.
We assume that the relaxation rate is much larger than the poisoning rate, i.e., $ \gamma_{\rm relax} \gg \gamma_{\rm qpp}$, corresponding to the reported measurements in recent experiments~\cite{vanLoo2025Single,Zhang2025Gate}.
As shown in Figs.~\ref{fig:G_Cq_pmm}(g)-~\ref{fig:G_Cq_pmm}(i), the averaged quantum capacitance shows a combination of resonance lines along both the $\mu_1=\mu_2$ and $\mu_1=-\mu_2$ directions, indicating a finite population of both $\ket{og}$ and $\ket{eg}$  as a consequence of parity switching due to the quasiparticle poisoning effect.
However, in contrast with the open regime, now distinction only exists in the magnitude of the quantum capacitance for different values of $\mu_A$.

\begin{figure}[tbp]
    \centering
    \includegraphics[width=\linewidth]{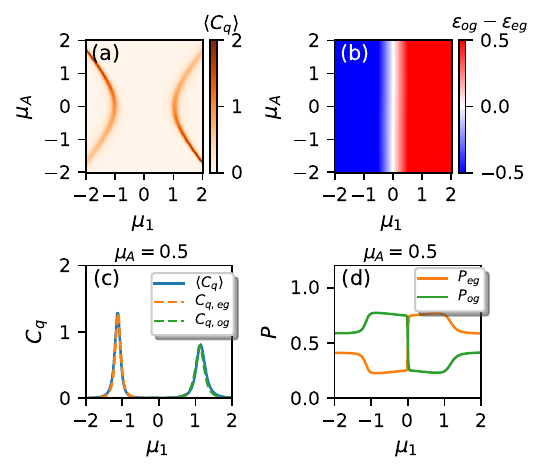}
    \caption{
    (a) Averaged quantum capacitance of a single quantum dot coupled to Andreev bound states in the $(\mu_1, \mu_A)$ plane.
    (b) Energy difference in the $(\mu_1, \mu_A)$ plane.
    (c) Solid blue curve: numerically calculated $\expval{C_q}$ as a function of $\mu_1$ for $\mu_A=0.5$. Dashed lines: analytically derived $C_{q,eg}$ and $C_{q,og}$ in Eqs.~\eqref{eq:Cq_DA_even} and~\eqref{eq:Cq_DA_odd}.
    Note that curves of $C_{q,eg}$ and $C_{q,og}$ are multiplied by a constant which is given by $P_{eg}(\mu_1=-E_A) = P_{og}(\mu_1=E_A) \approx 0.282$.
    (d) State population of $\ket{eg}$ and $\ket{og}$ as a function of $\mu_1$ for $\mu_A=0.5$.
    Parameters: $t_{\rm sc}=0.1, t_{\rm sf}=0.03, \gamma_{\rm lead} =0.003, k_BT_{\rm lead}=0.005, \gamma_{\rm qpp}=0.003, k_BT_{\rm qpp}=2, \gamma_{\rm relax}=1$.}
    \label{fig:Cq_mu1_muA}
\end{figure}

\section{Single quantum dot coupled with Andreev bound states}\label{sec:DA}

We now consider a single quantum dot coupled to Andreev bound states by detuning QD2 far from resonance.
A quantum dot-ABS pair is the minimal unit of a Kitaev chain and being able to probe the coupling between them is important for tuning Kitaev chains to desired regimes.
In particular, we focus on the nearly closed regime where both $\gamma_{\rm lead}$ and $\gamma_{\rm qpp}$ play an important role in determining the parity dynamics in the system.
Figure~\ref{fig:Cq_mu1_muA}(a) shows the averaged quantum capacitance in the $(\mu_1, \mu_A)$ plane.
A notable feature is the emergence of peaks at $\mu_1 = \pm \sqrt{\mu^2_A + \Delta^2_0}$, corresponding to energy alignment of the quantum dot and the Andreev bound states.
Moreover, the height and width of the peaks vary with the chemical potential of the ABS.
To gain deeper insight, we consider the low-energy effective Hamiltonian.
In the even-parity subspace, the effective Hamiltonian is given by
\begin{align}
    H_{\text{even}} = 
    \begin{pmatrix}
        0 & t_{\text{sc}} v  & -t_{\text{sf}} v  \\
        t_{\text{sc}} v  & \mu_1 + E_A & 0 \\
        -t_{\text{sf}} v & 0 & \mu_1 + E_A
    \end{pmatrix}
    \label{eq:H_even_DA}
\end{align}
under the basis of $\{ \ket{000}, \ket{110}, \ket{101} \}$ where $\ket{n_{1\sd}, n_{A\su},n_{A\sd}}$ denotes the occupancy of the quantum dot with spin down and Andreev bound states. 
Here, assuming strong Zeeman spin splitting, we neglect the spin-up orbitals in quantum dots.
Furthermore, in the weak coupling regime $t_{\rm sc}, t_{\rm sf} \ll \Delta_0$, $\ket{011}$ with an energy of $2E_A$ is also neglected.
It is thus straightforward to obtain the eigenstates, which we label as $\ket{eg},\ket{e1}$, and $\ket{e2}$.
In particular, the quantum capacitance of the ground state $\ket{eg}$ is 
\begin{align}
    C_{q,eg}=-\frac{d^2\varepsilon_{eg}}{d\mu^2_1} = \frac{t^2v^2}{4 \left[\left(\frac{\mu_1 + E_A}{2} \right)^2 + t^2v^2 \right]^{3/2}},
    \label{eq:Cq_DA_even}
\end{align}
where $t^2 = t^2_{\rm sc} + t^2_{\rm sf}$, and $u^2 = 1-v^2 = \frac{1}{2} + \frac{\mu_A}{2E_A} $ are the BCS coherence factors. 
The analytical results show a good agreement with the numerical calculations. 
In particular, $C_{q,eg}$ in Eq.~\eqref{eq:Cq_DA_even} describes the emergence of the peak at $\mu_1 = -E_A$ in Fig.~\ref{fig:Cq_mu1_muA}(a). 
Moreover, the $tv$ dependence in Eq.~\eqref{eq:Cq_DA_even} accounts for the narrowing of the peak as $\mu_A$ increases, and furthermore the strength of $tv$ can be extracted from the peak width.

Similarly, in the odd-parity subspace, the effective Hamiltonian is
\begin{align}
        H_{\text{odd}} = 
    \begin{pmatrix}
        \mu_1 & -t_{\text{sf}} u  & -t_{\text{sc}} u  \\
        -t_{\text{sf}} u  &  E_A & 0 \\
        -t_{\text{sc}} u & 0 & E_A
    \end{pmatrix}
    \label{eq:H_odd_DA}
\end{align}
under the basis of $\{ \ket{100}, \ket{010}, \ket{001} \}$.
The three eigenstates are $\ket{og}, \ket{o1},\ket{o2}$, with the quantum capacitance of $\ket{og}$ being
\begin{align}
    C_{q,og} = \frac{t^2u^2}{4 \left[\left(\frac{\mu_1 - E_A}{2} \right)^2 + t^2u^2 \right]^{3/2}}.
    \label{eq:Cq_DA_odd}
\end{align}
Now $C_{q,og}$ in  Eq.~\eqref{eq:Cq_DA_odd} corresponds to the peak centered at $\mu_1=E_A$ in Fig.~\ref{fig:Cq_mu1_muA}(a). 
However, by comparing Eq.~\eqref{eq:Cq_DA_even} and Fig.~\ref{fig:Cq_mu1_muA}(b), we also notice that the peak of $C_{q,eg}$ is located in a region where $\ket{og}$ is the global ground state, and that the eigenenergy of $\ket{eg}$ is higher by an amount of $\varepsilon_{eg} - \varepsilon_{og} \gtrsim \Delta_0$.
Since $C_{q,og}$ in Eq.~\eqref{eq:Cq_DA_odd} is essentially zero for $\mu_1 < 0$, the observation of a finite $\expval{C_q}$ around $\mu_1 = -E_A$ indicates that the nonequilibrium distribution gives rise to a finite $P_{eg}$ owing to quasiparticle poisoning.
Therefore, it sheds light on the energy scale appearing in the energy dependence of Eq.~\eqref{eq:Gamma_qpp}, i.e., $k_BT_{\rm qpp} \gtrsim \Delta_0$.
As a concrete example, we show a cutline of $\expval{C_q}$ along with $P_{eg}, P_{og}$ for $\mu_A=0.5$ [see Figs.~\ref{fig:Cq_mu1_muA}(c) and ~\ref{fig:Cq_mu1_muA}(d)].
By multiplying the analytical results of $C_{q,eg}$ and $C_{q,og}$ with a constant that depends on $P$ in Fig.~\ref{fig:Cq_mu1_muA}(d), we obtain an excellent agreement between the numerical and analytic results, see Fig.~\ref{fig:Cq_mu1_muA}(c).

\begin{figure}[tbp]
    \centering
    \includegraphics[width=\linewidth]{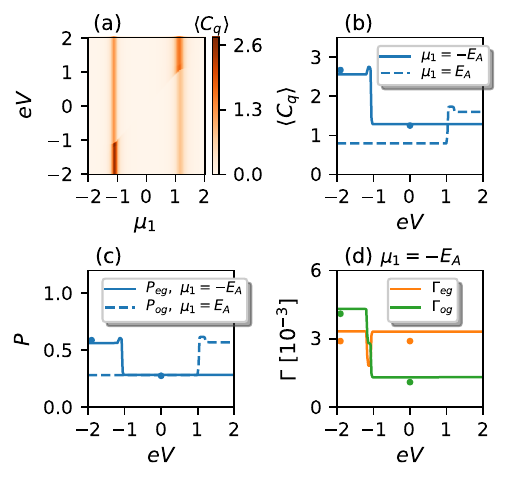}
    \caption{(a) $\langle C_q \rangle$ of a single quantum dot coupled to Andreev bound states in the $(\mu_1, eV)$ plane for $\mu_A=0.5$. 
    (b) Cutlines of $\langle C_q \rangle$ for $\mu_1 = \pm E_A \approx \pm 1.12$.
    (c) State population as a function of voltage bias.
    (d) Switching rate of $\ket{eg}$ and $\ket{og}$ as a function of voltage bias.
    Solid dots in panels (b), (c), and (d) are based on the analytical results obtained in Eqs.~\eqref{eq:Gamma_og}-\eqref{eq:Cq_analytic_Vneg}
    Parameters: $t_{\rm sc}=0.1, t_{\rm sf}=0.03, \gamma_{\rm lead} =0.003, k_BT_{\rm lead}=0.005, \gamma_{\rm qpp}=0.003, k_BT_{\rm qpp}=2, \gamma_{\rm relax}=1$.}
    \label{fig:Cq_mu1_mulead}
\end{figure}

To further reveal the combined effects of coupling to external normal leads and intrinsic quasiparticle poisoning on parity dynamics, we investigate the voltage dependence of the quantum capacitance as well as the switching rate.
Figure~\ref{fig:Cq_mu1_mulead}(a) shows $\expval{C_q}$ in the $(\mu_1, eV)$ plane for $\mu_A=0.5$.
A key feature is the bias-dependent enhancement of the two peaks: the peak at $\mu_1=-E_A$ increases for $eV < -E_A$, while the one at $\mu_1=E_A$ increases for $eV > E_A$, see Fig.~\ref{fig:Cq_mu1_mulead}(b).
For $\mu_1 = -E_A$, the first three eigenstates in the even-parity subspace are nearly degenerate with $\varepsilon_{eg}, \varepsilon_{e1}, \varepsilon_{e2} \approx O(t)$, see Eq.~\eqref{eq:H_even_DA}, while the energies of the odd-parity states are $\varepsilon_{og}\approx -E_A$, and $\varepsilon_{o1}, \varepsilon_{o2} \approx E_A$, see Eq.~\eqref{eq:H_odd_DA}.
At low bias $eV \approx 0$, the coupling to the external normal lead only allows transition from $\ket{eg}$ to $\ket{og}$, while the reverse direction is blocked.
Thus, the parity switching rate of $\ket{og}$ arises entirely from quasiparticle poisoning and is given by
\begin{align}
    \Gamma_{og} \equiv \sum_{\beta=eg,e1,e2}\Gamma_{\beta, og} \approx 3 \gamma_{\rm{qpp}}n_F(E_A, T_{\rm{qpp}}),
    \label{eq:Gamma_og}
\end{align} 
where a factor of 3 comes from the nearly three-fold degeneracy of the even states.
The switching rate of $\ket{eg}$ arises from both the external lead and quasiparticle poisoning and is given by
\begin{align}
    \Gamma_{eg} &\equiv \sum_{\beta=og,o1,o2}\Gamma_{\beta, eg} \nn
    & \approx \gamma_{\rm{lead}}/2 + \gamma_{\rm{qpp}} \left[1 + n_F(E_A, T_{\rm{qpp}})  \right],
\end{align}
where a factor of $1/2$ comes from the matrix element of $\abs{\bra{og} c\dg_{1\sd} \ket{eg}}^2 \approx 1/2$.
Therefore, the averaged quantum capacitance at zero bias is 
\begin{align}
\expval{C_q}_{V=0} \approx P_{eg} C_{q,eg} \approx \frac{\Gamma_{og}}{\Gamma_{og} + \Gamma_{eg}} C_{q,eg},
    \label{eq:Cq_analytic_V0}
\end{align}
where $C_{q,eg}$ is given by Eq.~\eqref{eq:Cq_DA_even}.
When $eV < -E_A - k_BT_{\rm{lead}}$, $\Gamma_{og}$ is enhanced because electron tunneling to the lead becomes allowed, while the switching rate due to quasiparticle poisoning remains the same.
Thus it becomes 
\begin{align}
    \Gamma'_{og} = \gamma_{\rm{lead}} + 3 \gamma_{\rm{qpp}}n_F(E_A, T_{\rm{qpp}}).
\end{align}
Here, the relevant nonzero matrix elements responsible for the finite contribution $\gamma_{\rm{lead}}$ are $\abs{\bra{eg} c_{1\sd} \ket{og}}^2 \approx \abs{\bra{e2} c_{1\sd} \ket{og}}^2 \approx 1/2$.
On the other hand, $\Gamma_{eg}$ at negative bias remains the same as $V=0$. 
Therefore, we obtain  
\begin{align}
    \expval{C_q}_{eV < -E_A } \approx \frac{\Gamma'_{og}}{\Gamma'_{og} + \Gamma_{eg}} C_{q,eg}.
    \label{eq:Cq_analytic_Vneg}
\end{align}
We thereby show that for $\mu_1 = -E_A$, $\expval{C_q}$ at negative bias ($V<E_A$) is larger than that at zero bias ($V=0$), consistent with numerical findings.
The analytically derived quantum capacitance, state population, and switching rates for $V=0$ and $V=-2$ are plotted as solid dots in Figs.~\ref{fig:Cq_mu1_mulead}(b), ~\ref{fig:Cq_mu1_mulead}(c), and~\ref{fig:Cq_mu1_mulead}(d), in good agreement with the exact numerical calculations.

\section{Discussion}
We consider a two-site Kitaev chain weakly coupled to an external normal lead and show that, in the open regime, quantum capacitance enables identification of the sweet spot, consistent with tunnel spectroscopy measurements. 
The key mechanism is that coupling to an electron reservoir relaxes the system to the global ground state, thereby allowing the quantum capacitance to reflect the ground-state parity.
On the other hand, we also consider the quantum capacitance of a single quantum dot coupled to Andreev bound states by detuning one of the quantum dots.
We propose a method to estimate several important Hamiltonian parameters, such as the dot-ABS coupling strength of $t_{\rm sc}$ and $t_{\rm sf}$, and the BCS coherence factors of the ABS.

Additionally, since quantum information of a Majorana qubit is encoded in the fermion parity of the ground-state manifold, it is crucial to understand the relevant mechanisms for parity switching between opposite parity subspaces, such as intrinsic quasiparticle poisoning and residual coupling to external normal leads.
Here, we show that a scan versus the lead voltage bias can provide additional insight into the parity dynamics of the system.
In particular, we propose how to estimate the switching rates of the parity ground states by comparing the quantum capacitance at zero and large bias.

The modeling of quasiparticle poisoning and relaxation processes in this work is phenomenological, however, it already captures several key features observed in recent experiments on Kitaev chain devices Refs.~\cite{vanLoo2025Single,Zhang2025Gate}.
We emphasize that our analysis and conclusions remain valid even if some details of the modeling are changed.
For example, the precise form of the poisoning function [e.g., $n_F(E,T_{\rm qpp})$ in Eq.~\eqref{eq:Gamma_qpp}] is inessential, while the key requirements are that it is a decreasing function of energy and that its energy dependence be smooth on a scale comparable to or larger than $\Delta_0$.
Similarly, alternative modeling of the relaxation processes would lead to qualitatively similar results, provided that it ensures sufficiantly fast relaxation to the ground state within the same parity subspace. 
We leave the studies of a more microscopic modeling of the poisoning and relaxation processes in future works.


\section{Summary}
To summarize, we show that quantum capacitance in the open regime can help to identify the sweet spot of a two-site Kitaev chain. 
For a single quantum dot coupled with Andreev bound states, the bias-dependent quantum capacitance can further reveal the parity dynamics of the system.
Our work provides helpful insights into the ongoing quantum capacitance experiments on quantum-dot-based Kitaev chain devices.

\begin{acknowledgments}
We acknowledge useful discussions with Yining Zhang, Nick van Loo, and Francesco Zatelli.
This work was supported by startup funds from Shanghai Jiao Tong University.
\end{acknowledgments}

\bibliography{references_CXL.bib}

\end{document}